\documentclass[aps,prd,twocolumn,groupedaddress,showpacs,
a4paper,floats]{revtex4}
\usepackage{graphicx}
\usepackage{amssymb,latexsym}
\usepackage{longtable}

\def\ga{\mathrel{\raise.3ex\hbox{$>$\kern-.75em\lower1ex\hbox{$\sim$}}}}
\def\la{\mathrel{\raise.3ex\hbox{$<$\kern-.75em\lower1ex\hbox{$\sim$}}}}
\def\be{\begin{equation}}
\def\ee{\end{equation}}
\def\ba{\begin{eqnarray}}
\def\ea{\end{eqnarray}}
\def\ga{\mathrel{\raise.3ex\hbox{$>$\kern-.75em\lower1ex\hbox{$\sim$}}}}
\def\la{\mathrel{\raise.3ex\hbox{$<$\kern-.75em\lower1ex\hbox{$\sim$}}}}

\begin{document}

\title{Reconstructing the dark energy equation of state with varying alpha}

\author{N. J. Nunes and James E. Lidsey}
%\thanks{}
\affiliation{Astronomy Unit, 
School of Mathematical Sciences, Queen Mary, University of
 London, Mile End Road, E1 4NS, UK}

\date{\today}

\begin{abstract}
The possibility of reconstructing the dark energy equation 
of state from variations in the fine structure constant 
is investigated for a class of models where the quintessence field 
is non--minimally coupled to the electromagnetic field. 
For given classes of couplings and quintessence interaction 
potentials, it is typically found that variations in alpha 
would need to be measured to within an accuracy of at 
least $5 \times 10^{-7}$ to obtain a reconstructed equation 
of state with less than a twenty per cent deviation from the true equation 
of state between redshifts 0 and 3. 
In this case, it is argued that the sign of the first 
derivative of the  equation of state can be uncovered from the 
reconstruction, thus providing unique 
information on how the universe developed into its present dark energy 
dominated phase independent of high redshift surveys.  
Such information would  
complement future observations anticipated from 
the Supernova Acceleration Probe.

\end{abstract}


\maketitle

%%%%%%%%%%%%%%%%%%%%%%%%%%%%%%%%%%%%%%%%%%%%%%%%%%%%%%%
\section{Introduction}
%%%%%%%%%%%%%%%%%%%%%%%%%%%%%%%%%%%%%%%%%%%%%%%%%%%%%%%

Some recent observations of a number of quasar absorption lines indicate that
the fine structure constant, $\alpha \equiv e^2/\hbar c$,
was smaller than its present value by 
$\Delta \alpha / \alpha = - 10^{-5}$ at redshifts 
in the range $z \sim
1-3$  
\cite{Webb:2000mn,Murphy:2003hw}. 
(See, however, Ref.~\cite{Chand:2004ct,Srianand:2004mq} 
for an independent 
analysis that does not support such a large variation in $\alpha$.)
Since this redshift range coincides with the epoch 
when the universe underwent a transition from matter domination to 
dark energy domination \cite{Riess:1998cb,Perlmutter:1998np}, 
it is of interest to consider the possibility that  
this change in the effective fine structure
constant arises as a direct result of a non-trivial 
gauge coupling between the dark energy and the 
electromagnetic field strength 
\cite{Carroll:1998bd,Dvali:2001dd,Olive:2001vz,Banks:2001qc,Chiba:2001er,Wetterich:2002ic,Wetterich:2003jt,Anchordoqui:2003ij,Parkinson:2003kf,Bertolami:2003qs,Lee:2003bg,Avelino:2004hu,Bento:2004jg} 
(see also \cite{Sandvik:2001rv,Kostelecky:2002ca,Barrow:2001iw,Barrow:2002ed,Barrow:2002hi,Mota:2003tc,Mota:2003tm,Khoury:2003rn,Vagenas:2003uk}).

In this paper we consider classes 
of models where the dark energy in the 
universe is identified as a slowly varying, self-interacting, neutral 
scalar ``quintessence'' field 
\cite{Wetterich:1988fm,Ratra:1988rm,Peebles:1988ek,Ferreira:1997au,Steinhardt:1999nw}
(see also the reviews \cite{Sahni:2002kh,Padmanabhan:2002ji})
that is minimally coupled to 
Einstein gravity but non-minimally coupled to 
the electromagnetic field. The action is given by  
\begin{eqnarray}
\label{action}
S = &-& \frac{1}{2 \kappa^2} \int d^4x \sqrt{-g}~ R \nonumber \\
    &+&
   \int d^4x \sqrt{-g} ~({\cal L}_{\phi} + {\cal L}_M + {\cal L}_{\phi F}) \,,
\end{eqnarray}
where $R$ is the Ricci curvature scalar of the metric $g_{\mu\nu}$, 
$g \equiv {\rm det}g_{\mu\nu}$, $\kappa^2 \equiv 
8\pi m_P^{-2}$ and $m_P$ is the Planck mass.
The Lagrangian density for the quintessence field is
\begin{equation}
{\cal L}_{\phi} = \frac{1}{2} \partial^{\mu}\phi\partial_{\mu}\phi
- V(\phi) \, ,
\end{equation}
where $V(\phi )$ is the self--interaction 
potential of the scalar field, $\phi$.
The interaction term between the scalar field and the
electromagnetic field is determined by
\begin{equation}
{\cal L}_{\phi F} = - \frac{1}{4} B_F(\phi) F_{\mu\nu}F^{\mu\nu} \,,
\label{bc}
\end{equation}
where 
$F_{\mu\nu}$ is the electromagnetic 
field strength and $B_F(\phi )$ 
represents the gauge kinetic function that parametrizes 
the coupling between the scalar and 
vector degrees of freedom. ${\cal L}_M$ represents the Lagrangian
density for the ordinary matter fields and we assume this sector to 
be dominated by a barotropic pressureless fluid.

The action (\ref{action}) is characterized in 
terms of two undetermined functions -- 
the gauge kinetic function and the self--interaction potential. 
In general, these would be determined by the nature of the 
underlying particle physics theory. For example,  
a generic exponential coupling of the 
type given in Eq.~(\ref{bc}) 
was introduced by Bekenstein \cite{Bekenstein:1982eu}. 
Exponential couplings between form--fields and scalar fields 
also arise generically in compactifications of string/M--theory 
to four dimensions, where the scalar field parametrizes the volume of the 
extra dimensions. (See, e.g., Ref.~\cite{Uzan:2002vq} for a recent 
review on the 
theoretical motivation of varying fundamental constants.)  

The gauge kinetic function specifies the value of   
the effective fine structure constant  
such that 
$\alpha = \alpha_0/B_F(\phi)$, where a subscript `0'
denotes the present--day value. 
The potential of the field is
related to the dark energy equation of state, $w_{\phi} \equiv 
p_{\phi}/\rho_{\phi}$, where $p_{\phi}^{} = \dot{\phi}^2/2 - V (\phi )$ 
represents the
pressure of the field,  
$\rho_{\phi}^{} = \dot{\phi}^2/2 + V (\phi )$ is the energy density and 
a dot denotes differentiation with respect to cosmic time.

A number of different approaches may be adopted when 
fitting models of the form (\ref{action}) 
to the data. In principle, the functions 
$\{ V(\phi ) , B_F(\phi ) \}$ would be determined 
within the context of a unified 
theory of the fundamental interactions, such as 
string/M--theory. In this case, a direct approach would be to determine 
the region of parameter space consistent with observations once these 
two functions have been specified. This  
approach was effectively followed recently by 
Parkinson, Bassett and Barrow \cite{Parkinson:2003kf}, 
who calculated the best fit parameters of a model
for which the exact form of the gauge kinetic function and the
dark energy equation of state were assumed {\em a priori}. 

We adopt an alternative approach in the present work by considering 
whether the quintessence potential or the gauge 
kinetic function can be {\em reconstructed} directly 
from observational data involving variations in the 
fine structure constant and the dark energy equation of state. 
For a given $w (z)$ and 
$\Delta \alpha /\alpha \equiv (\alpha -\alpha_0 )/\alpha_0$, 
there always exists a $B_F (\phi )$ that would fit the data.
Thus, a sufficiently accurate empirical 
determination of the equation of state, together with the 
evolution of $\alpha$, would allow 
the gauge kinetic function to be reconstructed. 

On the other hand, 
if the gauge kinetic function alone is 
specified {\em a priori} (either through theoretical
or phenomenological considerations), $w(z)$ and $\Delta \alpha /\alpha$ 
may no longer be viewed as independent variables, 
since they  
share a common origin through 
the rolling of the quintessence field. 
In effect, a consistency relation exists between 
these two quantities and an observational determination of one  
would constrain the other. 
This implies that the study of the 
absorption lines in quasar spectra 
can in principle yield additional information on 
variations in the dark energy equation of state 
(and the corresponding quintessence potential). 
This is important, given that a determination 
of the redshift dependence of the 
dark energy equation of state directly from the luminosity distance
relations is difficult -- the latter is determined by a double integral over 
the former and this can severely restrict the available 
information on the equation of state that can be extracted from observations
\cite{Maor:2000jy,Padmanabhan:2002vv}.

In this paper, we consider a linear dependence of the gauge
kinetic function on the scalar field: 
\begin{equation}
\label{defBF}
B_F(\phi) = 1-\zeta\kappa (\phi-\phi_0)
\end{equation}
where $\zeta$ 
is a constant. This dependence may be viewed as arising from a Taylor 
expansion of a generic gauge kinetic function
and is expected to be valid for a wide class of models when 
$\kappa(\phi-\phi_0) < 1$ is satisfied
over the range of redshifts
relevant to observations, $z \approx 0 - 4$. 
It then follows that  
the effective fine structure constant depends on the value of 
the quintessence field such that 
\cite{Olive:2001vz,Copeland:2003cv,Anchordoqui:2003ij}
\begin{equation}
\label{gaugekfunc}
\frac{\Delta \alpha}{\alpha} \equiv \frac{\alpha-\alpha_0}{\alpha_0} =
\zeta \kappa (\phi - \phi_0) \,.
\end{equation}

Assuming that the mass of the scalar 
field effectively vanishes, 
tests of the equivalence principle imply that the parameter, $\zeta$,
is bounded by $| \zeta |< 10^{-3}$ \cite{Olive:2001vz}. 
Bounds on variations in the fine structure constant arise from 
the Oklo natural nuclear reactor 
($|\Delta \alpha/\alpha | < 10^{-7}$ at redshift $z = 0.14$
\cite{Damour:1996zw,Olive:2002tz}
\footnote{However one should
emphasize that the analysis performed in
Ref.~\cite{Lamoreaux:2003ii} for the Oklo natural reactor 
suggests a larger $\alpha$ than today's 
with $\Delta \alpha/ \alpha \geq 4.5 \times 10^{-8}$.}),
and the meteorite constraint
($|\Delta \alpha/\alpha| < 10^{-6}$ at redshift $z = 0.45$
\cite{Olive:2003sq}). 

For consistency, neither the constraint 
arising from the Oklo natural nuclear reactor 
nor the meteorite constraint
were considered in this work, although 
it is generally expected that significant 
variations should be observed up to a redshift of order unity
in quintessence models. However, the model in Fig.~\ref{AS}, 
naturally satisfies the former bound at redshift $z=0.14$.
As discussed in Ref.~\cite{Copeland:2003cv}, 
these bounds can
be evaded through the existence of a form factor in the coupling
$\zeta$ with respect to the photon momentum. Such a form factor 
can result in changes in $\alpha$ at the level of atomic physics
without leading to observable effects on nuclear phenomena. 
Moreover, it has been shown for a particular model
in Ref.~\cite{Mota:2003tm}, that if dark
energy collapses along with dark matter this would naturally lead to a
significant difference between the value of the fine structure constant in our
galaxy and the one in the background.
For these reasons, we decide to explore also the models that do not
satisfy the Oklo and meteorites bounds at low redshifts.

%%%%%%%%%%%%%%%%%%%%%%%%%%%%%%%%%%%%%%%%%%%%%%%%%%%%%%%%%%%%%%%%%%
\section{Reconstructing the equation of state: in principle}
%%%%%%%%%%%%%%%%%%%%%%%%%%%%%%%%%%%%%%%%%%%%%%%%%%%%%%%%%%%%%%%%%%

To proceed we consider a spatially homogeneous 
quintessence field propagating in the spatially flat 
Friedmann-Robertson-Walker (FRW) universe. We 
assume that the contribution of the electromagnetic degrees 
of freedom to the total energy density of the universe is negligible
and consequently that the cosmic dynamics is determined by the 
scalar field and a background pressureless fluid (corresponding to 
dark and visible matter).
It then follows that the cosmic dynamics is 
determined by the Einstein equation 
\begin{equation}
\label{dotH}
H H' = -\frac{\kappa^2}{2}\left(\rho_M + H^2 \phi'^2 \right) \,, \\
%\label{dusteom}
%\rho_M' = -3\rho_M \,
\end{equation}
and scalar field equation
\begin{equation}
\label{scalareom}
\rho_{\phi}' = -3H^2 \phi'^2 \,,
\end{equation}
subject to the Friedmann constraint
\begin{equation}
\label{friedmann}
H^2 = \frac{\kappa^2}{3} (\rho_M + \rho_{\phi}) \,,
\end{equation}
where a prime denotes differentiation with respect to $N = \ln a$, 
$H \equiv \dot{a}/a$ is the Hubble expansion parameter,  
the matter density is given by $\rho_M=\rho_0 {\Omega_M}_0/a^3$
and $\rho_0$ denotes the present value of the critical 
energy density. 
The term $(dB_F(\phi)/d\phi ~F_{\mu\nu}F^{\mu\nu})$ containing 
the derivative of the gauge kinetic function was neglected in 
Eq.~(\ref{scalareom}) as its statistical average over the present 
Hubble radius is zero for photons \cite{Olive:2001vz}.

Only two of the equations (\ref{dotH})--(\ref{friedmann}) 
are independent and the cosmic dynamics is fully determined once the 
functional form of the quintessence potential, $V(\phi )$, has been specified. 
The nature of the potential determines how the field evolves in time
and the corresponding variations 
in the fine structure constant 
are then determined by Eq.~(\ref{gaugekfunc}). 
Moreover, the equation of state is defined in terms of the 
field's kinetic and potential energies:  
\begin{equation}
\label{eosdef}
w \equiv \frac{\dot{\phi}^2-2V(\phi)}{\dot{\phi}^2+2V(\phi)}
\end{equation}
and its dependence on redshift also follows given the form of the potential. 

The principle idea of the inversion procedure is that variations in the fine 
structure constant may be employed within the context of this 
class of models to deduce changes in the kinetic energy 
of the scalar field at higher redshifts. Eqs.~(\ref{dotH})--(\ref{scalareom})
may then be employed to determine the corresponding changes in the field's
energy density, or equivalently, its potential, and hence the equation of 
state from Eq.~(\ref{eosdef}). This procedure is analogous to that employed
in analysing 
the classical dynamics of a particle, $x$, moving in a one-dimensional
potential well. 
The form of the well determines the particle's motion, 
and this can be represented as a trajectory in the phase space parametrized by 
$\{ x ,\dot{x} \}$. Equivalently, the corresponding 
potential can be reconstructed once the appropriate trajectory has been 
specified.

We now develop the inversion procedure. 
Substituting
Eq.~(\ref{friedmann}) into
Eq.~(\ref{scalareom}) yields a differential equation for the
evolution of the energy density of the scalar field:
\begin{eqnarray}
\label{doU}
\sigma' = -(\kappa\phi')^2 (\sigma + a^{-3}) 
\,,
\end{eqnarray}
where we have defined $\sigma = \rho_{\phi}/\rho_0 {\Omega_M}_0$. 
The general solution to Eq.~(\ref{doU}) can be expressed 
in terms of quadratures with respect to the kinetic 
energy of the quintessence field: 
\begin{eqnarray}
\label{gensol}
\sigma (N) &=& e^{- \int^N_0 dN (\kappa \phi')^2} \times
\nonumber \\ 
&~& \times \left[ 
\sigma_0  -  \int^N_0 dN (\kappa \phi')^2 
e^{ -3N + \int^N_0 dN (\kappa \phi')^2}
\right] \nonumber \,, \\
\end{eqnarray}
where the integration constant $\sigma_0$ is defined such that 
$\sigma(N=0) \equiv \sigma_0 = {\Omega_{\phi}}_0/{\Omega_M}_0$. 

The dark energy equation of state is given 
in general by $w= -1-(\ln \sigma )'/3$ and 
substitution of Eq.~(\ref{doU}) implies that 
\begin{equation}
\label{eosphi}
w(N) = -1 + \frac{(\kappa \phi')^2}{3} \left( 1+
\frac{1}{\sigma a^3} \right) \,.
\end{equation}
Hence, the equation of state can be reconstructed 
once the redshift dependence of the first derivative of the field has been 
determined. It is important to emphasize that 
only the first derivative needs to be measured. If the gauge kinetic function 
is known, this dependence can be inferred directly from variations in 
the fine structure constant.

It follows from the definition $\dot{\phi}^2 
= \rho_{\phi}+p_{\phi}$ that the equation of state (\ref{eosphi}) 
can also be expressed in the form  
\begin{equation}
\label{wOmega}
w=-1+\frac{(\kappa\phi')^2}{3\Omega_{\phi}} \,,
\end{equation}
where $\Omega_{\phi} \equiv 
\kappa^2 \rho_{\phi}/(3H^2)$. 
Differentiating Eq.~(\ref{wOmega}) with respect to scale 
and substituting 
Eqs.~(\ref{dotH}), (\ref{scalareom}) and (\ref{friedmann})
then implies that the first derivative (running) 
of the equation of state can be 
expressed directly in terms of first two derivatives of the scalar 
field: 
\begin{equation}
\label{wprime}
w' = 2(1+w) \frac{\phi''}{\phi'} +3 w \left( 1+w -
\frac{\left( \kappa \phi' \right)^2}{3} \right) \,,
\end{equation}
where $w' = -(dw/dz)/a$. An infinite hierarchy of expressions relating 
the $n$--th derivative of the equation of state 
to the $(n+1)$--th derivative of the field 
could be derived. Each represents a consistency relation 
between the equation of state and variations in the fine structure 
constant once the 
gauge kinetic function has been specified. 

We now illustrate the above reconstruction 
procedure with a specific example 
where $\kappa(\phi-\phi_0) = \lambda N$ for some constant 
$\lambda$.  
In this case, the integrals in Eq.~(\ref{gensol}) can 
be evaluated analytically: 
\begin{equation}
\label{gensolconstant}
\sigma = \left( \frac{{\Omega_{\phi}}_0}{{\Omega_M}_0} 
+\frac{\lambda^2}{\lambda^2 -3}
\right) e^{-\lambda^2 N} - \left( 
\frac{\lambda^2}{\lambda^2 -3} \right) e^{-3N} \,,
\end{equation}
where $\lambda^2 = 3 {\Omega_{\phi}}_0 (1+w_0)$.  
The equation of state is then deduced by substituting 
Eq.~(\ref{gensolconstant}) into Eq.~(\ref{eosphi}): 
\begin{equation}
w(N) = (\lambda^2-3)\left[3- \frac{\lambda^2}{w_0} 
\frac{{\Omega_M}_0}{{\Omega_{\phi}}_0} 
\exp \left((\lambda^2-3)N\right)\right]^{-1} \,.
\end{equation}
Finally, 
the quintessence potential can be reconstructed by noting that 
$V(N) = {\Omega_M}_0 \rho_0 \sigma -\phi'^2H^2/2$ 
and employing the Friedmann equation (\ref{friedmann}). 
We find that
\begin{equation}
V= Ae^{-\frac{3}{\lambda}\kappa \phi} - B e^{-\lambda \kappa \phi } \,,
\end{equation}
where the mass scales $A$ and $B$ are positive--definite and given by
\begin{eqnarray}
A &=& \frac{1}{2}~ \frac{\lambda^2}{3-\lambda^2} ~\rho_0 {\Omega_M}_0 ~
e^{\frac{3}{\lambda}\kappa \phi_0} \,, \\
B &=& \frac{1}{2} ~\frac{6-\lambda^2}{3-\lambda^2} ~
\rho_0 {\Omega_{\phi}}_0 w_0 ~e^{\lambda\kappa \phi_0} \,,
\end{eqnarray} 
respectively.
In the above example, it was assumed implicitly that 
the gauge kinetic function was such that   
the cosmological variation of $\Delta \alpha/\alpha$ corresponded 
to a variation in the scalar field of the form $\phi \propto
N$. The form of $B_F(\phi)$ was not specified. 

%%%%%%%%%%%%%%%%%%%%%%%%%%%%%%%%%%%%%%%%%%%%%%%%%%%%%%%%%%%%%%%%%%
\section{Reconstructing the equation of state: in practice}
%%%%%%%%%%%%%%%%%%%%%%%%%%%%%%%%%%%%%%%%%%%%%%%%%%%%%%%%%%%%%%%%%%
In this section we consider the reconstruction of three 
dark energy equations of state by employing the method 
outlined in the previous section for a gauge kinetic function 
given by Eq.~(\ref{defBF}). 
The scalar field potentials have been 
investigated previously within the context 
of viable quintessence models 
\cite{Brax:1999gp,Barreiro:1999zs,Albrecht:1999rm,Anchordoqui:2003ij}. 
The reconstructions
are shown in Figs.~\ref{sugra} -- \ref{AS}. These examples
correspond to three different possible evolutions for $w(z)$, 
namely, those cases where it increases, decreases or
oscillates with increasing redshift. The second and third examples
are particularly important as they can not be reproduced with the 
parametrization employed in Ref.~\cite{Parkinson:2003kf}, 
since in that work
the equation of state is always increasing with increasing redshift.
For the models studied in the present work, 
we have verified that $\kappa (\phi - \phi_0 )<1$ over the appropriate 
range of redshifts, and this is consistent with the interpretation of 
Eq.~(\ref{defBF}) as a lowest-order Taylor expansion of a 
generic gauge kinetic function. Let us now describe the reconstruction
process. 

\subsection{Generating simulated data}
In testing the reconstruction procedure, it is necessary to first
generate simulated data sets 
for the variations in $\alpha$. 
This was achieved by specifying the functional 
form of the quintessence potential and numerically 
integrating the field equations (\ref{dotH})--(\ref{friedmann}) to
determine the redshift dependence of both the 
equation of state and the quintessence field. 
The latter determines the corresponding variations in $\alpha$ 
from Eq.~(\ref{gaugekfunc}) once the coupling parameter, 
$\zeta$, has been specified.
We then generated the simulated data set, with 
associated error bars, for 
$\Delta \alpha /\alpha$ based on the exact numerical solution. 
Specifically, the data points are equally spaced 
in the redshift range $z \in [ 0.2, 4]$ at intervals of $0.2$
and are normally
distributed with mean $\zeta \kappa (\phi -\phi_0 )$. 
In each example, 
the value of $\zeta$ was chosen so that the variations 
in $\zeta \kappa (\phi -\phi_0 )$ resulted in changes in the 
fine structure constant of the order $\Delta \alpha /\alpha \approx 
10^{-5}$, as observed in 
the present QSO data \cite{Murphy:2003hw}.

\subsection{Fitting the data}
The reconstruction can then proceed by 
fitting the generated data points to
a polynomial function 
\begin{equation}
\label{gdef}
g(N) = \frac{\Delta \alpha}{\alpha} = g_1 N + g_2 N^2 + ... \,,
\end{equation}
where $g_i$ are constants. The result of these fits is shown in
Fig.~\ref{simdata}. Equations (\ref{gaugekfunc}) and 
(\ref{gdef}) map the variations 
in $\alpha$ as measured
by QSO observations onto the corresponding variations 
in the scalar field for a given value of the coupling constant, 
$\zeta$.  
The degree of the polynomial (\ref{gdef}) 
employed in the fitting differs for the 
three cases because the underlying equations of state exhibit different 
levels of complexity. For the class of models we have studied we found
that a polynomial of degree three provides generally a good fit to
the generated data. However, for models with an 
oscillating equation of state, one
needs to increase the degree of the polynomial in order to obtain 
both a successful reconstruction and a 
reduced $\chi^2$ of order unity. 

\subsection{Estimating $\zeta$}
The first and second derivatives of
the field are related to $g$ such that 
$\phi' = g'/\zeta\kappa$ and 
$\phi'' = g''/\zeta\kappa$, respectively.
In practice, therefore,  
the numerical value of the coupling parameter must be estimated 
empirically since the reconstruction via Eqs.~(\ref{gensol}) 
and (\ref{eosphi}) requires the scale dependence of the 
quintessence field to be known. Substituting 
Eq.~(\ref{gdef}) into Eq.~(\ref{wOmega}) implies that 
\begin{equation}
\label{zeta1}
\zeta^2 = \frac{1}{3}~\frac{g'^2}{\Omega_{\phi}(1+w)} \,,
\end{equation}
and it follows from 
Eq. (\ref{zeta1}) that a numerical estimate for the
coupling may be
deduced given the present-day values of the quintessence 
field's energy density 
and the equation of state $w_0$, 
together with the variation, $g_0'$, in the fine structure constant, as 
determined from QSO observations. 
For example, given the typical values ${\Omega_{\phi}}_0 \sim 0.7$, 
$-w_0 \sim
0.6 - 0.99$, and $g_0' \sim 10^{-7} - 10^{-5}$, 
we find that $\zeta \sim 10^{-7} -
10^{-4}$, in accordance with the values obtained in 
Ref.~\cite{Copeland:2003cv}, where
specific quintessence models were studied. This range of values of
$\zeta$ is compatible with bounds  arising from tests of the equivalence
principle which demand $|\zeta| < 10^{-3}$ \cite{Olive:2001vz}.

However, one source of uncertainty in the reconstruction procedure 
is the uncertainty in the present--day 
value of the equation of state, $w_0$. The latest
measurements constrain this parameter within the range 
$-1.38< w_0 <-0.82$ at the 95$\%$ confidence level, assuming a constant 
equation of state
\cite{Melchiorri:2002ux}. This uncertainty generates an 
uncertainty in the value of the coupling $\zeta$ and, in view of this,  
we allowed the equation of state to take a range of possible 
values, $\tilde{w}_0$. 
More specifically, in Figs.~\ref{sugra} and \ref{2exp}, the 
present value of the equation of state was chosen to be 
$\tilde{w}_0 = w_0 + (-0.1,0,0.1)$ 
where $w_0$ represents the
correct value as deduced from the numerical integration.
In Fig.~\ref{AS}, on the other hand, the 
present value was chosen to be $\tilde{w}_0 = w_0 +
(1+w_0)(0.9,0,-0.95)$, respectively, 
when moving downward in the figure.  

It follows from 
Eq.~(\ref{zeta1}) that, when $w \approx -1$, a small uncertainty in
the value of $\tilde{w}_0$ can lead to a large uncertainty in 
the value of the coupling constant $\zeta$ and
consequently to distinct possible evolutions for the corresponding 
equation of state. A typical case of $w
\approx -1$ arises when the scalar field 
undergoes oscillations about
the minimum of its potential. On the other hand, 
an expression equivalent to Eq.~(\ref{wprime}) for the form
of $B_F(\phi)$ adopted in this paper is given by 
\begin{equation}
\label{zeta2}
\zeta^2 =
\frac{g'^2}{w'}~\frac{\Omega_M}{\Omega_{\phi}}~
\left(w+\frac{2}{3}\frac{1}{\Omega_M} \frac{g''}{g'} \right) \,.
\end{equation}
It follows, therefore, that in such cases 
a more accurate estimate 
of the magnitude of $\zeta$ can be made 
if information on the present--day value of the 
first derivative of the equation
of state is also available. We note that the quantity $dw/dz$ is an observable 
parameter believed to be within reach of future SnIa observations from 
the Supernova Acceleration Probe (SNAP)
\cite{Perlmutter:2003kf}. 

We must note, however, that if $\zeta$ was known on
fundamental particle physics grounds, the full reconstruction of the
equation of state could be achieved without any need to normalize it
to an independent result.

\subsection{Reconstruction results}
\begin{figure}
\includegraphics[width=8cm]{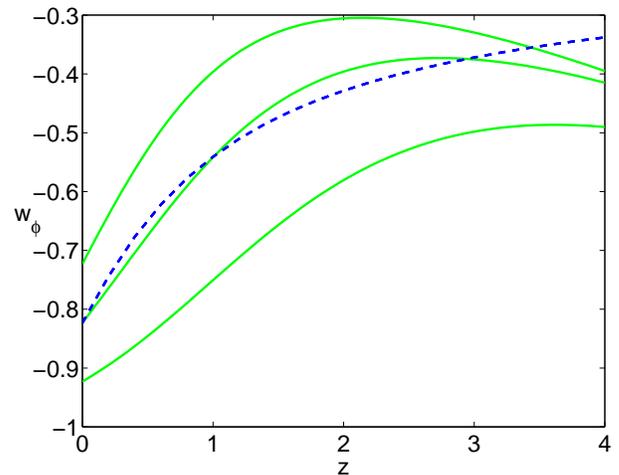}
\caption{\label{sugra} 
Dashed line: evolution of the equation of state of the scalar
field $\phi$ with potential $V(\phi) = V_0 \exp( (\kappa \phi)^2/2)/\phi^{11}$ 
determined from numerical integration of the field equations 
\cite{Brax:1999gp}. 
Solid lines: evolution of the reconstructed equation of
state for different possible values of the present--day equation
of state (see the text for details). 
The function $g(N)$ was fitted as a polynomial of degree 3 
to a set of points normally
distributed with a standard deviation of $1 \times 10^{-7}$.}
\end{figure}

\begin{figure}
\includegraphics[width=8cm]{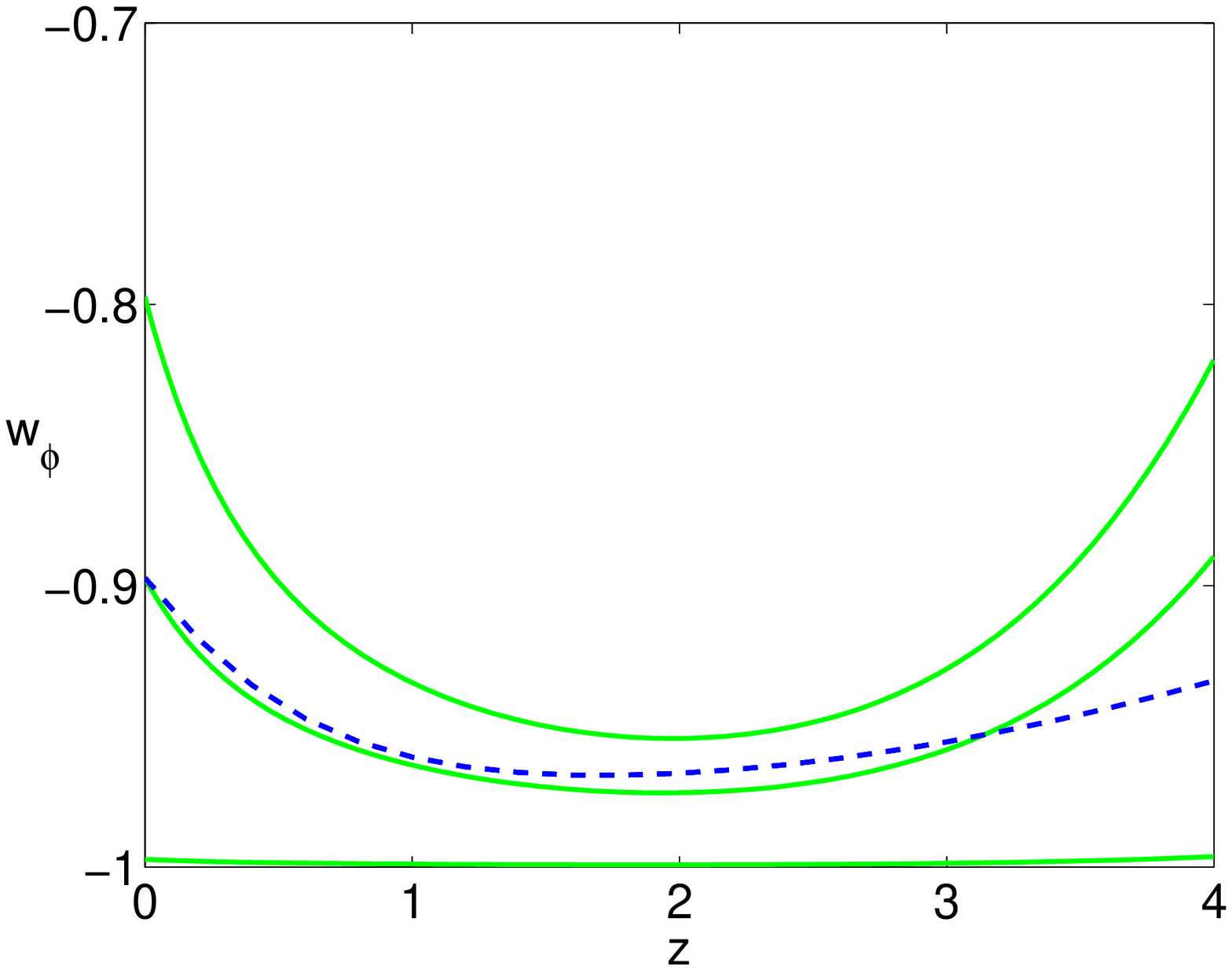}
\caption{\label{2exp} 
Dashed line: evolution of the equation of state of the scalar
field $\phi$ with potential $V(\phi) = V_0 (\exp(50 \kappa \phi) +
\exp(0.8 \kappa \phi))$ determined from numerical 
integration of the field equations \cite{Barreiro:1999zs}. 
Solid lines: evolution of the reconstructed equation of
state for different possible values of the present--day equation
of state (see the text for details).
The function $g(N)$ was fitted as a polynomial of degree 3 
to a set of points normally
distributed with a standard deviation of $5 \times 10^{-7}$.}
\end{figure}

\begin{figure}
\includegraphics[width=8cm]{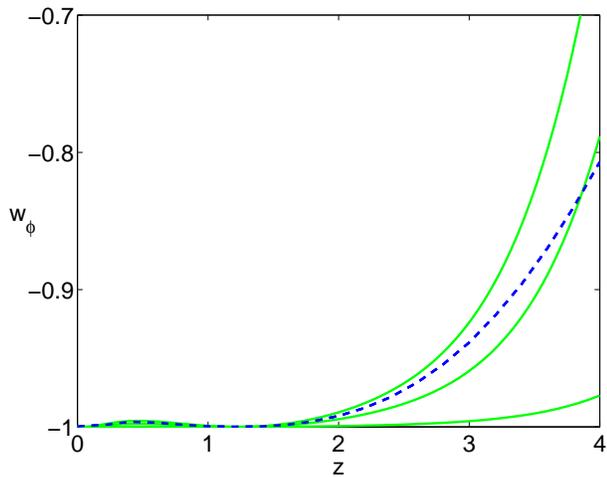}
\caption{\label{AS} 
Dashed line: evolution of the equation of state of the scalar
field $\phi$ with potential 
$V = \kappa^{-4} e^{-A \kappa \phi}\left[ (\kappa \phi - C)^2 +
B)\right]$ determined from numerical 
integration of the field equations, 
where $A = 8.5$ and $B = 0.93/A^2$
\cite{Albrecht:1999rm,Anchordoqui:2003ij}. 
Solid lines: evolution of the reconstructed equation of
state for different possible values of the present--day equation
of state (see the text for details).
The function $g(N)$ was fitted as a polynomial of degree 5 
to a set of points normally
distributed with a standard deviation of $5 \times 10^{-7}$.}
\end{figure}

The results of the reconstructions 
are illustrated in Figs.~\ref{sugra} -- \ref{AS}, 
where Eq.~(\ref{zeta1}) has been employed to estimate the
value of the coupling $\zeta$ for the different choices of $\tilde{w}_0$ 
by substituting $w \rightarrow \tilde{w}_0$ and $g'
\rightarrow g_1$. In each case,
the dashed line in the figures
represents the exact numerical solution of the equations of motion
when ${\Omega_{\phi}}_0 = 0.7$. The corresponding 
solid lines illustrate the reconstructed evolution for the different
values of $\tilde{w}_0$ considered. 
An uncertainty in the present value of the equation
of state of $\delta w_0 \approx 0.1$ is expected from the SNAP
data. Hence, the lines with $\tilde{w}_0 = w_0 \pm 0.1$ (in Figs.~\ref{sugra}
and \ref{2exp}) 
define the error band on the evolution of $w$ arising from the uncertainty we
will have on $w_0$. It is worth emphasizing that SNAP will
provide data within the range of redshifts between 0 and 1.7, whereas
the QSO data can (in principle) provide us with information on the equation of
state out to a redshift as high as $z = 4$.

\begin{figure}
\includegraphics[width=8cm]{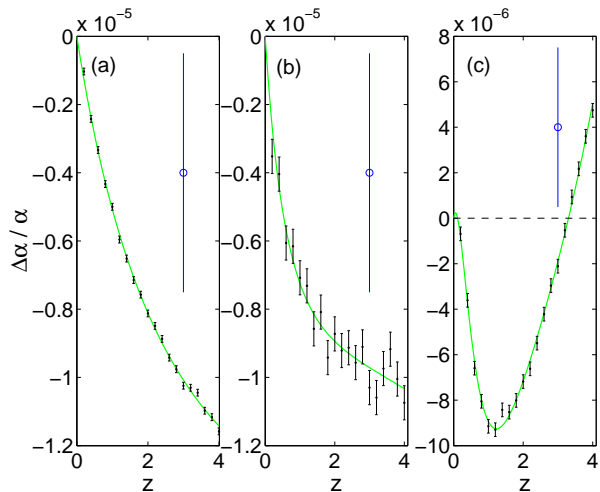}
\caption{\label{simdata}
Evolution of $\Delta \alpha/\alpha$ for the three quintessence models
studied in this work. The error bars represent the maximum 
uncertainty allowed in order to obtain a reliable reconstruction of
the equation of state with no more than $20\%$ deviation from that 
deduced from the numerical integration.
(a): $V(\phi) = V_0 \exp( (\kappa \phi)^2/2)/\phi^{11}$ 
\cite{Brax:1999gp}; 
(b): $V(\phi) = V_0 (\exp(50 \kappa \phi) +
\exp(0.8 \kappa \phi))$ \cite{Barreiro:1999zs}; 
(c): $V = \kappa^{-4} e^{-A \kappa \phi}\left[ (\kappa \phi - C)^2 +
B)\right]$, 
where $A = 8.5$ and $B = 0.93/A^2$
\cite{Albrecht:1999rm,Anchordoqui:2003ij}. The large error bar in the
figures represent a typical current uncertainty on $\Delta \alpha/\alpha$
from QSO observations of $\delta(\Delta \alpha/\alpha) \approx 3.5
\times 10^{-6}$.}
\end{figure}

When investigating the sensitivity of the reconstruction procedure
on errors in the variations of $\alpha$, we typically found that 
in order to obtain a reconstruction, $\tilde{w}(z)$, with no more
than a $20\%$ deviation from the true equation of state, $w(z)$, i.e.,
\begin{equation}
\frac{\tilde{w}(z)-w(z)}{w(z)} < 0.2 \,,
\end{equation}
(for $\tilde{w}_0 = w_0$), one requires an observational 
determination of $\Delta \alpha/\alpha$ to within an accuracy of 
{\em at least} $\sim 5 \times 10^{-7}$ between redshifts 0 and 3. 
Figure \ref{simdata} shows how small the error bars on data
points should be in order to obtain a reliable reconstruction 
at this order. 
This is an order of magnitude 
smaller than the expected sensitivity of the  
High Accuracy Radial Velocity Planet Searcher spectrograph
$\delta(\Delta \alpha/\alpha) \approx 10^{-6}$.
\cite{Levshakov:2003fa}.

%%%%%%%%%%%%%%%%%%%%%%%%%%%%%%%%%%%%%%%%%%%
\section{Discussion}
%%%%%%%%%%%%%%%%%%%%%%%%%%%%%%%%%%%%%%%%%%%
Cosmological observations including high redshift surveys 
of type Ia supernovae and the anisotropy power spectrum of 
the Cosmic Microwave Background (CMB) indicate that the present-day 
value of the dark energy equation of state is bounded by
$-1.38< w_0 <-0.82$ at the 95$\%$ confidence level, assuming a constant 
equation of state \cite{Melchiorri:2002ux}. Such bounds would be 
weakened for a wider class of models where the equation of state is 
allowed to vary, but at present  
there are only very weak observational constraints on the 
``running'' of the equation of state, $d w/dz =-aw'$
\cite{Riess:2004nr}, and it 
is not yet possible to distinguish such models from a cosmological constant.
In this paper, we have investigated the possibility 
that further information on the redshift dependence 
of the equation of state  
can be deduced independently of high redshift surveys through observed 
variations in the fine structure constant. In principle, the 
reconstruction of the equation of state is possible if the 
form of the gauge kinetic function that couples the scalar and 
electromagnetic fields is known. The advantage of a
reconstruction of this type is that it yields information on the equation 
of state at redshifts significantly higher than the limited range 
accessible to SNAP (corresponding to $z \le 1.7$).

The primary question addressed in the present paper is how
much information one could acquire on $w(z)$ from variations 
in the fine structure constant alone.
In a full reconstruction,
one would employ all the data available, from both supernovae surveys and
measurements of $\Delta \alpha /\alpha$, and perform a 
full cross analysis between the different data sets. 
However, one must also establish
what can be learned from each data set independently. Indeed, this is a
necessary and crucial step in the program we have outlined,
precisely because $w(z)$ and $\Delta \alpha /\alpha$ are not
independent as they share a common origin through the quintessence 
potential. 
As a result, information on variations in the 
equation of state determined separately from supernova surveys (see, e.g. 
\cite{Nakamura:1998mt,Saini:1999ba,Chiba:2000im,Alam:2003sc}) and 
quasar surveys (as presented above) 
should be consistent. Establishing an inconsistency
would indicate that a reliable reconstruction could
not be achieved and, furthermore, would immediately rule
out this class of models, namely the form of $B_F(\phi)$, 
as a mechanism for
correlating dark energy and variations in $\alpha$.

Figs.~\ref{sugra} -- \ref{AS}
indicate that the reconstructions do yield information 
on whether the running of the equation of state is positive or negative, 
at least out to a redshift $z \approx 3$. Although the 
error in the normalization of $w_0$ implies that the uncertainties in the 
magnitude of the reconstructed first derivative may be large, 
the qualitative shape of the equation of state can be deduced,
provided variations in $\alpha$ are determined to within an
accuracy of $5 \times 10^{-7}$ or better. We have 
performed equivalent analyses for the potentials considered in Section III 
over different regions of parameter space, as well as for other quintessence 
potentials, and have arrived at similar conclusions. 

Any information that can be extracted directly from observations
on whether the equation of state increases or decreases with redshift 
is of importance. For example,  
if the equation of state 
increases with redshift (i.e. $w$ moves away from $-1$),  
this implies that the field is slowing down as we approach the present day.  
On the other hand,  the 
kinetic energy of the field is growing as 
the universe expands if
$w$ decreases with increasing redshift. 
This latter behaviour could correspond, for
example, to a creeping quintessence scenario \cite{Steinhardt:1999nw}, 
where the field 
has overshot the attractor value and has started to move only very
recently. Thus, information on the 
first derivative of the equation of state provides us with 
unique insight into how the universe 
underwent the transition from matter domination to dark energy domination. 

For the general class of models defined by Eqs. (\ref{action}) 
and (\ref{gaugekfunc}), 
the qualitative behaviour of the equation of state 
can be deduced directly from Eq.~(\ref{zeta1})
without the need to solve Eq.~(\ref{doU}) if it is observed 
that $g'^2$ increases with redshift. 
Since Eq.~(\ref{zeta1}) is valid over all scales and $\Omega_{\phi}$ is a
decreasing function of redshift, it necessarily follows that 
the equation of state must have been larger
in the past and this case would therefore rule out the 
possibility of a creeping quintessence scenario. On the other hand, 
for the case where $g'^2$ is a decreasing variable,
we must proceed to solve Eq.~(\ref{doU}) directly in order to gain further 
insight. 

Finally, we outline a complementary approach that 
may allow the equation of state and its derivatives 
to be deduced at a specific redshift. This approach corresponds 
to a perturbative reconstruction 
of the equation of state. It follows from Eqs. (\ref{wprime}) 
and (\ref{gaugekfunc}) 
that the first derivative of the equation of state at a given redshift can be 
directly determined if the corresponding values of $\{ w , g' , g'' \}$
are known. Higher derivatives can also be constrained if sufficient 
information on the corresponding derivatives of the fitting function 
$g(N)$ is available. Assuming that the necessary constraints on the 
derivatives 
could be determined from QSO observations, the one remaining 
free parameter would be the equation of state, or equivalently 
from Eq. (\ref{wOmega}), the density of the dark energy.  
This parameter could in turn be deduced from the Friedmann 
equation (\ref{friedmann}) 
if the Hubble parameter, $H(z)$, were known and  
this could be found from the luminosity distance, $d_L$: 
\begin{equation}
H^{-1} (z) = \frac{d}{dz} \left( \frac{d_L}{1+z} \right)
\end{equation}
It would be interesting to explore this possibility further.

\begin{acknowledgments}
NJN is supported by the Particle Physics and 
Astronomy Research Council (PPARC). We thank C. Martins
and M. Pospelov for helpful comments and suggestions.
\end{acknowledgments}

%\bibliography{../Biblio/alpha.bib,../Biblio/parameters.bib,../Biblio/mypapers.bib,../Biblio/dynsyst.bib,../Biblio/qmodels.bib,../Biblio/wobs.bib,../Biblio/wprmz.bib,../Biblio/cc.bib}

\begin{thebibliography}{55}
\expandafter\ifx\csname natexlab\endcsname\relax\def\natexlab#1{#1}\fi
\expandafter\ifx\csname bibnamefont\endcsname\relax
  \def\bibnamefont#1{#1}\fi
\expandafter\ifx\csname bibfnamefont\endcsname\relax
  \def\bibfnamefont#1{#1}\fi
\expandafter\ifx\csname citenamefont\endcsname\relax
  \def\citenamefont#1{#1}\fi
\expandafter\ifx\csname url\endcsname\relax
  \def\url#1{\texttt{#1}}\fi
\expandafter\ifx\csname urlprefix\endcsname\relax\def\urlprefix{URL }\fi
\providecommand{\bibinfo}[2]{#2}
\providecommand{\eprint}[2][]{\url{#2}}

\bibitem[{\citenamefont{Webb et~al.}(2001)}]{Webb:2000mn}
\bibinfo{author}{\bibfnamefont{J.~K.} \bibnamefont{Webb}} \bibnamefont{et~al.},
  \bibinfo{journal}{Phys. Rev. Lett.} \textbf{\bibinfo{volume}{87}},
  \bibinfo{pages}{091301} (\bibinfo{year}{2001}), \eprint{astro-ph/0012539}.

\bibitem[{\citenamefont{Murphy et~al.}(2003)\citenamefont{Murphy, Webb, and
  Flambaum}}]{Murphy:2003hw}
\bibinfo{author}{\bibfnamefont{M.~T.} \bibnamefont{Murphy}},
  \bibinfo{author}{\bibfnamefont{J.~K.} \bibnamefont{Webb}}, \bibnamefont{and}
  \bibinfo{author}{\bibfnamefont{V.~V.} \bibnamefont{Flambaum}}
  (\bibinfo{year}{2003}), \eprint{astro-ph/0306483}.

\bibitem[{\citenamefont{Chand et~al.}(2004)\citenamefont{Chand, Srianand,
  Petitjean, and Aracil}}]{Chand:2004ct}
\bibinfo{author}{\bibfnamefont{H.}~\bibnamefont{Chand}},
  \bibinfo{author}{\bibfnamefont{R.}~\bibnamefont{Srianand}},
  \bibinfo{author}{\bibfnamefont{P.}~\bibnamefont{Petitjean}},
  \bibnamefont{and} \bibinfo{author}{\bibfnamefont{B.}~\bibnamefont{Aracil}}
  (\bibinfo{year}{2004}), \eprint{astro-ph/0401094}.

\bibitem[{\citenamefont{Srianand et~al.}(2004)\citenamefont{Srianand, Chand,
  Petitjean, and Aracil}}]{Srianand:2004mq}
\bibinfo{author}{\bibfnamefont{R.}~\bibnamefont{Srianand}},
  \bibinfo{author}{\bibfnamefont{H.}~\bibnamefont{Chand}},
  \bibinfo{author}{\bibfnamefont{P.}~\bibnamefont{Petitjean}},
  \bibnamefont{and} \bibinfo{author}{\bibfnamefont{B.}~\bibnamefont{Aracil}}
  (\bibinfo{year}{2004}), \eprint{astro-ph/0402177}.

\bibitem[{\citenamefont{Riess et~al.}(1998)}]{Riess:1998cb}
\bibinfo{author}{\bibfnamefont{A.~G.} \bibnamefont{Riess}} \bibnamefont{et~al.}
  (\bibinfo{collaboration}{Supernova Search Team}), \bibinfo{journal}{Astron.
  J.} \textbf{\bibinfo{volume}{116}}, \bibinfo{pages}{1009}
  (\bibinfo{year}{1998}), \eprint[http://arXiv.org/abs]{astro-ph/9805201}.

\bibitem[{\citenamefont{Perlmutter et~al.}(1999)}]{Perlmutter:1998np}
\bibinfo{author}{\bibfnamefont{S.}~\bibnamefont{Perlmutter}}
  \bibnamefont{et~al.} (\bibinfo{collaboration}{Supernova Cosmology Project}),
  \bibinfo{journal}{Astrophys. J.} \textbf{\bibinfo{volume}{517}},
  \bibinfo{pages}{565} (\bibinfo{year}{1999}), \eprint{astro-ph/9812133}.

\bibitem[{\citenamefont{Carroll}(1999)}]{Carroll:1998bd}
\bibinfo{author}{\bibfnamefont{S.~M.} \bibnamefont{Carroll}},
  \bibinfo{journal}{AIP Conf. Proc.} \textbf{\bibinfo{volume}{478}},
  \bibinfo{pages}{291} (\bibinfo{year}{1999}), \eprint{astro-ph/9806099}.

\bibitem[{\citenamefont{Dvali and Zaldarriaga}(2002)}]{Dvali:2001dd}
\bibinfo{author}{\bibfnamefont{G.~R.} \bibnamefont{Dvali}} \bibnamefont{and}
  \bibinfo{author}{\bibfnamefont{M.}~\bibnamefont{Zaldarriaga}},
  \bibinfo{journal}{Phys. Rev. Lett.} \textbf{\bibinfo{volume}{88}},
  \bibinfo{pages}{091303} (\bibinfo{year}{2002}), \eprint{hep-ph/0108217}.

\bibitem[{\citenamefont{Olive and Pospelov}(2002)}]{Olive:2001vz}
\bibinfo{author}{\bibfnamefont{K.~A.} \bibnamefont{Olive}} \bibnamefont{and}
  \bibinfo{author}{\bibfnamefont{M.}~\bibnamefont{Pospelov}},
  \bibinfo{journal}{Phys. Rev. D} \textbf{\bibinfo{volume}{65}},
  \bibinfo{pages}{085044} (\bibinfo{year}{2002}), \eprint{hep-ph/0110377}.

\bibitem[{\citenamefont{Banks et~al.}(2002)\citenamefont{Banks, Dine, and
  Douglas}}]{Banks:2001qc}
\bibinfo{author}{\bibfnamefont{T.}~\bibnamefont{Banks}},
  \bibinfo{author}{\bibfnamefont{M.}~\bibnamefont{Dine}}, \bibnamefont{and}
  \bibinfo{author}{\bibfnamefont{M.~R.} \bibnamefont{Douglas}},
  \bibinfo{journal}{Phys. Rev. Lett.} \textbf{\bibinfo{volume}{88}},
  \bibinfo{pages}{131301} (\bibinfo{year}{2002}), \eprint{hep-ph/0112059}.

\bibitem[{\citenamefont{Chiba and Kohri}(2002)}]{Chiba:2001er}
\bibinfo{author}{\bibfnamefont{T.}~\bibnamefont{Chiba}} \bibnamefont{and}
  \bibinfo{author}{\bibfnamefont{K.}~\bibnamefont{Kohri}},
  \bibinfo{journal}{Prog. Theor. Phys.} \textbf{\bibinfo{volume}{107}},
  \bibinfo{pages}{631} (\bibinfo{year}{2002}), \eprint{hep-ph/0111086}.

\bibitem[{\citenamefont{Wetterich}(2002)}]{Wetterich:2002ic}
\bibinfo{author}{\bibfnamefont{C.}~\bibnamefont{Wetterich}}
  (\bibinfo{year}{2002}), \eprint{hep-ph/0203266}.

\bibitem[{\citenamefont{Wetterich}(2003)}]{Wetterich:2003jt}
\bibinfo{author}{\bibfnamefont{C.}~\bibnamefont{Wetterich}},
  \bibinfo{journal}{Phys. Lett. B} \textbf{\bibinfo{volume}{561}},
  \bibinfo{pages}{10} (\bibinfo{year}{2003}), \eprint{hep-ph/0301261}.

\bibitem[{\citenamefont{Anchordoqui and Goldberg}(2003)}]{Anchordoqui:2003ij}
\bibinfo{author}{\bibfnamefont{L.}~\bibnamefont{Anchordoqui}} \bibnamefont{and}
  \bibinfo{author}{\bibfnamefont{H.}~\bibnamefont{Goldberg}}
  (\bibinfo{year}{2003}), \eprint{hep-ph/0306084}.

\bibitem[{\citenamefont{Parkinson et~al.}(2003)\citenamefont{Parkinson,
  Bassett, and Barrow}}]{Parkinson:2003kf}
\bibinfo{author}{\bibfnamefont{D.}~\bibnamefont{Parkinson}},
  \bibinfo{author}{\bibfnamefont{B.~A.} \bibnamefont{Bassett}},
  \bibnamefont{and} \bibinfo{author}{\bibfnamefont{J.~D.} \bibnamefont{Barrow}}
  (\bibinfo{year}{2003}), \eprint{astro-ph/0307227}.

\bibitem[{\citenamefont{Bertolami et~al.}(2003)\citenamefont{Bertolami,
  Lehnert, Potting, and Ribeiro}}]{Bertolami:2003qs}
\bibinfo{author}{\bibfnamefont{O.}~\bibnamefont{Bertolami}},
  \bibinfo{author}{\bibfnamefont{R.}~\bibnamefont{Lehnert}},
  \bibinfo{author}{\bibfnamefont{R.}~\bibnamefont{Potting}}, \bibnamefont{and}
  \bibinfo{author}{\bibfnamefont{A.}~\bibnamefont{Ribeiro}}
  (\bibinfo{year}{2003}), \eprint{astro-ph/0310344}.

\bibitem[{\citenamefont{Lee et~al.}(2003)\citenamefont{Lee, Lee, and
  Ng}}]{Lee:2003bg}
\bibinfo{author}{\bibfnamefont{D.-S.} \bibnamefont{Lee}},
  \bibinfo{author}{\bibfnamefont{W.}~\bibnamefont{Lee}}, \bibnamefont{and}
  \bibinfo{author}{\bibfnamefont{K.-W.} \bibnamefont{Ng}}
  (\bibinfo{year}{2003}), \eprint{astro-ph/0309316}.

\bibitem[{\citenamefont{Avelino et~al.}(2004)\citenamefont{Avelino, Martins,
  and Oliveira}}]{Avelino:2004hu}
\bibinfo{author}{\bibfnamefont{P.~P.} \bibnamefont{Avelino}},
  \bibinfo{author}{\bibfnamefont{C.~J. A.~P.} \bibnamefont{Martins}},
  \bibnamefont{and} \bibinfo{author}{\bibfnamefont{J.~C. R.~E.}
  \bibnamefont{Oliveira}} (\bibinfo{year}{2004}), \eprint{astro-ph/0402379}.

\bibitem[{\citenamefont{Bento et~al.}(2004)\citenamefont{Bento, Bertolami, and
  Santos}}]{Bento:2004jg}
\bibinfo{author}{\bibfnamefont{M.~d.~C.} \bibnamefont{Bento}},
  \bibinfo{author}{\bibfnamefont{O.}~\bibnamefont{Bertolami}},
  \bibnamefont{and} \bibinfo{author}{\bibfnamefont{N.~M.~C.}
  \bibnamefont{Santos}} (\bibinfo{year}{2004}), \eprint{astro-ph/0402159}.

\bibitem[{\citenamefont{Sandvik et~al.}(2002)\citenamefont{Sandvik, Barrow, and
  Magueijo}}]{Sandvik:2001rv}
\bibinfo{author}{\bibfnamefont{H.~B.} \bibnamefont{Sandvik}},
  \bibinfo{author}{\bibfnamefont{J.~D.} \bibnamefont{Barrow}},
  \bibnamefont{and} \bibinfo{author}{\bibfnamefont{J.}~\bibnamefont{Magueijo}},
  \bibinfo{journal}{Phys. Rev. Lett.} \textbf{\bibinfo{volume}{88}},
  \bibinfo{pages}{031302} (\bibinfo{year}{2002}), \eprint{astro-ph/0107512}.

\bibitem[{\citenamefont{Barrow et~al.}(2002{\natexlab{a}})\citenamefont{Barrow,
  Sandvik, and Magueijo}}]{Barrow:2001iw}
\bibinfo{author}{\bibfnamefont{J.~D.} \bibnamefont{Barrow}},
  \bibinfo{author}{\bibfnamefont{H.~B.} \bibnamefont{Sandvik}},
  \bibnamefont{and} \bibinfo{author}{\bibfnamefont{J.}~\bibnamefont{Magueijo}},
  \bibinfo{journal}{Phys. Rev. D} \textbf{\bibinfo{volume}{65}},
  \bibinfo{pages}{063504} (\bibinfo{year}{2002}{\natexlab{a}}),
  \eprint{astro-ph/0109414}.

\bibitem[{\citenamefont{Barrow and Mota}(2002)}]{Barrow:2002ed}
\bibinfo{author}{\bibfnamefont{J.~D.} \bibnamefont{Barrow}} \bibnamefont{and}
  \bibinfo{author}{\bibfnamefont{D.~F.} \bibnamefont{Mota}},
  \bibinfo{journal}{Class. Quant. Grav.} \textbf{\bibinfo{volume}{19}},
  \bibinfo{pages}{6197} (\bibinfo{year}{2002}), \eprint{gr-qc/0207012}.

\bibitem[{\citenamefont{Barrow et~al.}(2002{\natexlab{b}})\citenamefont{Barrow,
  Magueijo, and Sandvik}}]{Barrow:2002hi}
\bibinfo{author}{\bibfnamefont{J.~D.} \bibnamefont{Barrow}},
  \bibinfo{author}{\bibfnamefont{J.}~\bibnamefont{Magueijo}}, \bibnamefont{and}
  \bibinfo{author}{\bibfnamefont{H.~B.} \bibnamefont{Sandvik}},
  \bibinfo{journal}{Phys. Lett. B} \textbf{\bibinfo{volume}{541}},
  \bibinfo{pages}{201} (\bibinfo{year}{2002}{\natexlab{b}}),
  \eprint{astro-ph/0204357}.

\bibitem[{\citenamefont{Mota and Barrow}(2003{\natexlab{a}})}]{Mota:2003tc}
\bibinfo{author}{\bibfnamefont{D.~F.} \bibnamefont{Mota}} \bibnamefont{and}
  \bibinfo{author}{\bibfnamefont{J.~D.} \bibnamefont{Barrow}}
  (\bibinfo{year}{2003}{\natexlab{a}}), \eprint{astro-ph/0306047}.

\bibitem[{\citenamefont{Mota and Barrow}(2003{\natexlab{b}})}]{Mota:2003tm}
\bibinfo{author}{\bibfnamefont{D.~F.} \bibnamefont{Mota}} \bibnamefont{and}
  \bibinfo{author}{\bibfnamefont{J.~D.} \bibnamefont{Barrow}}
  (\bibinfo{year}{2003}{\natexlab{b}}), \eprint{astro-ph/0309273}.

\bibitem[{\citenamefont{Khoury and Weltman}(2003)}]{Khoury:2003rn}
\bibinfo{author}{\bibfnamefont{J.}~\bibnamefont{Khoury}} \bibnamefont{and}
  \bibinfo{author}{\bibfnamefont{A.}~\bibnamefont{Weltman}}
  (\bibinfo{year}{2003}), \eprint{astro-ph/0309411}.

\bibitem[{\citenamefont{Kostelecky et~al.}(2002)\citenamefont{Kostelecky,
  Lehnert, and Perry}}]{Kostelecky:2002ca}
\bibinfo{author}{\bibfnamefont{V.~A.} \bibnamefont{Kostelecky}},
  \bibinfo{author}{\bibfnamefont{R.}~\bibnamefont{Lehnert}}, \bibnamefont{and}
  \bibinfo{author}{\bibfnamefont{M.~J.} \bibnamefont{Perry}}
  (\bibinfo{year}{2002}), \eprint{astro-ph/0212003}.

\bibitem[{\citenamefont{Vagenas}(2003)}]{Vagenas:2003uk}
\bibinfo{author}{\bibfnamefont{E.~C.} \bibnamefont{Vagenas}},
  \bibinfo{journal}{JHEP} \textbf{\bibinfo{volume}{07}}, \bibinfo{pages}{046}
  (\bibinfo{year}{2003}), \eprint{hep-th/0307192}.

\bibitem[{\citenamefont{Wetterich}(1988)}]{Wetterich:1988fm}
\bibinfo{author}{\bibfnamefont{C.}~\bibnamefont{Wetterich}},
  \bibinfo{journal}{Nucl. Phys. B} \textbf{\bibinfo{volume}{302}},
  \bibinfo{pages}{668} (\bibinfo{year}{1988}).

\bibitem[{\citenamefont{Ratra and Peebles}(1988)}]{Ratra:1988rm}
\bibinfo{author}{\bibfnamefont{B.}~\bibnamefont{Ratra}} \bibnamefont{and}
  \bibinfo{author}{\bibfnamefont{P.~J.~E.} \bibnamefont{Peebles}},
  \bibinfo{journal}{Phys. Rev. D} \textbf{\bibinfo{volume}{37}},
  \bibinfo{pages}{3406} (\bibinfo{year}{1988}).

\bibitem[{\citenamefont{Peebles and Ratra}(1988)}]{Peebles:1988ek}
\bibinfo{author}{\bibfnamefont{P.~J.~E.} \bibnamefont{Peebles}}
  \bibnamefont{and} \bibinfo{author}{\bibfnamefont{B.}~\bibnamefont{Ratra}},
  \bibinfo{journal}{Astrophys. J.} \textbf{\bibinfo{volume}{325}},
  \bibinfo{pages}{L17} (\bibinfo{year}{1988}).

\bibitem[{\citenamefont{Ferreira and Joyce}(1997)}]{Ferreira:1997au}
\bibinfo{author}{\bibfnamefont{P.~G.} \bibnamefont{Ferreira}} \bibnamefont{and}
  \bibinfo{author}{\bibfnamefont{M.}~\bibnamefont{Joyce}},
  \bibinfo{journal}{Phys. Rev. Lett.} \textbf{\bibinfo{volume}{79}},
  \bibinfo{pages}{4740} (\bibinfo{year}{1997}),
  \eprint[http://arXiv.org/abs]{astro-ph/9707286}.

\bibitem[{\citenamefont{Steinhardt et~al.}(1999)\citenamefont{Steinhardt, Wang,
  and Zlatev}}]{Steinhardt:1999nw}
\bibinfo{author}{\bibfnamefont{P.~J.} \bibnamefont{Steinhardt}},
  \bibinfo{author}{\bibfnamefont{L.-M.} \bibnamefont{Wang}}, \bibnamefont{and}
  \bibinfo{author}{\bibfnamefont{I.}~\bibnamefont{Zlatev}},
  \bibinfo{journal}{Phys. Rev. D} \textbf{\bibinfo{volume}{59}},
  \bibinfo{pages}{123504} (\bibinfo{year}{1999}),
  \eprint[http://arXiv.org/abs]{astro-ph/9812313}.

\bibitem[{\citenamefont{Sahni}(2002)}]{Sahni:2002kh}
\bibinfo{author}{\bibfnamefont{V.}~\bibnamefont{Sahni}},
  \bibinfo{journal}{Class. Quant. Grav.} \textbf{\bibinfo{volume}{19}},
  \bibinfo{pages}{3435} (\bibinfo{year}{2002}),
  \eprint[http://arXiv.org/abs]{astro-ph/0202076}.

\bibitem[{\citenamefont{Padmanabhan}(2003)}]{Padmanabhan:2002ji}
\bibinfo{author}{\bibfnamefont{T.}~\bibnamefont{Padmanabhan}},
  \bibinfo{journal}{Phys. Rept.} \textbf{\bibinfo{volume}{380}},
  \bibinfo{pages}{235} (\bibinfo{year}{2003}), \eprint{hep-th/0212290}.

\bibitem[{\citenamefont{Bekenstein}(1982)}]{Bekenstein:1982eu}
\bibinfo{author}{\bibfnamefont{J.~D.} \bibnamefont{Bekenstein}},
  \bibinfo{journal}{Phys. Rev. D} \textbf{\bibinfo{volume}{25}},
  \bibinfo{pages}{1527} (\bibinfo{year}{1982}).

\bibitem[{\citenamefont{Uzan}(2003)}]{Uzan:2002vq}
\bibinfo{author}{\bibfnamefont{J.-P.} \bibnamefont{Uzan}},
  \bibinfo{journal}{Rev. Mod. Phys.} \textbf{\bibinfo{volume}{75}},
  \bibinfo{pages}{403} (\bibinfo{year}{2003}), \eprint{hep-ph/0205340}.

\bibitem[{\citenamefont{Maor et~al.}(2001)\citenamefont{Maor, Brustein, and
  Steinhardt}}]{Maor:2000jy}
\bibinfo{author}{\bibfnamefont{I.}~\bibnamefont{Maor}},
  \bibinfo{author}{\bibfnamefont{R.}~\bibnamefont{Brustein}}, \bibnamefont{and}
  \bibinfo{author}{\bibfnamefont{P.~J.} \bibnamefont{Steinhardt}},
  \bibinfo{journal}{Phys. Rev. Lett.} \textbf{\bibinfo{volume}{86}},
  \bibinfo{pages}{6} (\bibinfo{year}{2001}), \eprint{astro-ph/0007297}.

\bibitem[{\citenamefont{Padmanabhan and Choudhury}(2003)}]{Padmanabhan:2002vv}
\bibinfo{author}{\bibfnamefont{T.}~\bibnamefont{Padmanabhan}} \bibnamefont{and}
  \bibinfo{author}{\bibfnamefont{T.~R.} \bibnamefont{Choudhury}},
  \bibinfo{journal}{Mon. Not. Roy. Astron. Soc.}
  \textbf{\bibinfo{volume}{344}}, \bibinfo{pages}{823} (\bibinfo{year}{2003}),
  \eprint{astro-ph/0212573}.

\bibitem[{\citenamefont{Copeland et~al.}(2003)\citenamefont{Copeland, Nunes,
  and Pospelov}}]{Copeland:2003cv}
\bibinfo{author}{\bibfnamefont{E.~J.} \bibnamefont{Copeland}},
  \bibinfo{author}{\bibfnamefont{N.~J.} \bibnamefont{Nunes}}, \bibnamefont{and}
  \bibinfo{author}{\bibfnamefont{M.}~\bibnamefont{Pospelov}}
  (\bibinfo{year}{2003}), \eprint{hep-ph/0307299}.

\bibitem[{\citenamefont{Damour and Dyson}(1996)}]{Damour:1996zw}
\bibinfo{author}{\bibfnamefont{T.}~\bibnamefont{Damour}} \bibnamefont{and}
  \bibinfo{author}{\bibfnamefont{F.}~\bibnamefont{Dyson}},
  \bibinfo{journal}{Nucl. Phys. B} \textbf{\bibinfo{volume}{480}},
  \bibinfo{pages}{37} (\bibinfo{year}{1996}), \eprint{hep-ph/9606486}.

\bibitem[{\citenamefont{Olive et~al.}(2002)}]{Olive:2002tz}
\bibinfo{author}{\bibfnamefont{K.~A.} \bibnamefont{Olive}}
  \bibnamefont{et~al.}, \bibinfo{journal}{Phys. Rev. D}
  \textbf{\bibinfo{volume}{66}}, \bibinfo{pages}{045022}
  (\bibinfo{year}{2002}), \eprint{hep-ph/0205269}.

\bibitem[{\citenamefont{Olive et~al.}(2003)}]{Olive:2003sq}
\bibinfo{author}{\bibfnamefont{K.~A.} \bibnamefont{Olive}} \bibnamefont{et~al.}
  (\bibinfo{year}{2003}), \eprint{astro-ph/0309252}.

\bibitem[{\citenamefont{Brax and Martin}(1999)}]{Brax:1999gp}
\bibinfo{author}{\bibfnamefont{P.}~\bibnamefont{Brax}} \bibnamefont{and}
  \bibinfo{author}{\bibfnamefont{J.}~\bibnamefont{Martin}},
  \bibinfo{journal}{Phys. Lett. B} \textbf{\bibinfo{volume}{468}},
  \bibinfo{pages}{40} (\bibinfo{year}{1999}),
  \eprint[http://arXiv.org/abs]{astro-ph/9905040}.

\bibitem[{\citenamefont{Barreiro et~al.}(2000)\citenamefont{Barreiro, Copeland,
  and Nunes}}]{Barreiro:1999zs}
\bibinfo{author}{\bibfnamefont{T.}~\bibnamefont{Barreiro}},
  \bibinfo{author}{\bibfnamefont{E.~J.} \bibnamefont{Copeland}},
  \bibnamefont{and} \bibinfo{author}{\bibfnamefont{N.~J.} \bibnamefont{Nunes}},
  \bibinfo{journal}{Phys. Rev. D} \textbf{\bibinfo{volume}{61}},
  \bibinfo{pages}{127301} (\bibinfo{year}{2000}), \eprint{astro-ph/9910214}.

\bibitem[{\citenamefont{Albrecht and Skordis}(2000)}]{Albrecht:1999rm}
\bibinfo{author}{\bibfnamefont{A.}~\bibnamefont{Albrecht}} \bibnamefont{and}
  \bibinfo{author}{\bibfnamefont{C.}~\bibnamefont{Skordis}},
  \bibinfo{journal}{Phys. Rev. Lett.} \textbf{\bibinfo{volume}{84}},
  \bibinfo{pages}{2076} (\bibinfo{year}{2000}),
  \eprint[http://arXiv.org/abs]{astro-ph/9908085}.

\bibitem[{\citenamefont{Melchiorri et~al.}(2003)\citenamefont{Melchiorri,
  Mersini, Odman, and Trodden}}]{Melchiorri:2002ux}
\bibinfo{author}{\bibfnamefont{A.}~\bibnamefont{Melchiorri}},
  \bibinfo{author}{\bibfnamefont{L.}~\bibnamefont{Mersini}},
  \bibinfo{author}{\bibfnamefont{C.~J.} \bibnamefont{Odman}}, \bibnamefont{and}
  \bibinfo{author}{\bibfnamefont{M.}~\bibnamefont{Trodden}},
  \bibinfo{journal}{Phys. Rev. D} \textbf{\bibinfo{volume}{68}},
  \bibinfo{pages}{043509} (\bibinfo{year}{2003}), \eprint{astro-ph/0211522}.

\bibitem[{\citenamefont{Perlmutter and Schmidt}(2003)}]{Perlmutter:2003kf}
\bibinfo{author}{\bibfnamefont{S.}~\bibnamefont{Perlmutter}} \bibnamefont{and}
  \bibinfo{author}{\bibfnamefont{B.~P.} \bibnamefont{Schmidt}}
  (\bibinfo{year}{2003}), \eprint{astro-ph/0303428}.

\bibitem[{\citenamefont{Levshakov}(2003)}]{Levshakov:2003fa}
\bibinfo{author}{\bibfnamefont{S.~A.} \bibnamefont{Levshakov}}
  (\bibinfo{year}{2003}), \eprint{astro-ph/0309817}.

\bibitem[{\citenamefont{Riess et~al.}(2004)}]{Riess:2004nr}
\bibinfo{author}{\bibfnamefont{A.~G.} \bibnamefont{Riess}} \bibnamefont{et~al.}
  (\bibinfo{year}{2004}), \eprint{astro-ph/0402512}.

\bibitem[{\citenamefont{Nakamura and Chiba}(1999)}]{Nakamura:1998mt}
\bibinfo{author}{\bibfnamefont{T.}~\bibnamefont{Nakamura}} \bibnamefont{and}
  \bibinfo{author}{\bibfnamefont{T.}~\bibnamefont{Chiba}},
  \bibinfo{journal}{Mon. Not. Roy. Astron. Soc.}
  \textbf{\bibinfo{volume}{306}}, \bibinfo{pages}{696} (\bibinfo{year}{1999}),
  \eprint{astro-ph/9810447}.

\bibitem[{\citenamefont{Saini et~al.}(2000)\citenamefont{Saini, Raychaudhury,
  Sahni, and Starobinsky}}]{Saini:1999ba}
\bibinfo{author}{\bibfnamefont{T.~D.} \bibnamefont{Saini}},
  \bibinfo{author}{\bibfnamefont{S.}~\bibnamefont{Raychaudhury}},
  \bibinfo{author}{\bibfnamefont{V.}~\bibnamefont{Sahni}}, \bibnamefont{and}
  \bibinfo{author}{\bibfnamefont{A.~A.} \bibnamefont{Starobinsky}},
  \bibinfo{journal}{Phys. Rev. Lett.} \textbf{\bibinfo{volume}{85}},
  \bibinfo{pages}{1162} (\bibinfo{year}{2000}),
  \eprint[http://arXiv.org/abs]{astro-ph/9910231}.

\bibitem[{\citenamefont{Chiba and Nakamura}(2000)}]{Chiba:2000im}
\bibinfo{author}{\bibfnamefont{T.}~\bibnamefont{Chiba}} \bibnamefont{and}
  \bibinfo{author}{\bibfnamefont{T.}~\bibnamefont{Nakamura}},
  \bibinfo{journal}{Phys. Rev.} \textbf{\bibinfo{volume}{D62}},
  \bibinfo{pages}{121301} (\bibinfo{year}{2000}), \eprint{astro-ph/0008175}.

\bibitem[{\citenamefont{Alam et~al.}(2003)\citenamefont{Alam, Sahni, Saini, and
  Starobinsky}}]{Alam:2003sc}
\bibinfo{author}{\bibfnamefont{U.}~\bibnamefont{Alam}},
  \bibinfo{author}{\bibfnamefont{V.}~\bibnamefont{Sahni}},
  \bibinfo{author}{\bibfnamefont{T.~D.} \bibnamefont{Saini}}, \bibnamefont{and}
  \bibinfo{author}{\bibfnamefont{A.~A.} \bibnamefont{Starobinsky}}
  (\bibinfo{year}{2003}), \eprint{astro-ph/0303009}.

\bibitem[{\citenamefont{Lamoreaux}(2003)}]{Lamoreaux:2003ii}
\bibinfo{author}{\bibfnamefont{S.~K.} \bibnamefont{Lamoreaux}}
  (\bibinfo{year}{2003}), \eprint{nucl-th/0309048}.

\end{thebibliography}

\end{document}